\documentstyle[prd,aps,preprint,tighten,epsfig]{revtex}
\begin{document}
\draft
\title{Lepton Mass Hierarchy and Neutrino Mixing}
\author{{\bf Harald Fritzsch} $^a$ ~ and ~ {\bf Zhi-zhong Xing} $^b$
\thanks{E-mail: xingzz@mail.ihep.ac.cn}}
\address{$^a$ Sektion Physik, Universit$\it\ddot{a}$t M$\it\ddot{u}$nchen,
Theresienstrasse 37A, 80333 Munich, Germany \\
$^b$ Institute of High Energy Physics, Chinese Academy of
Sciences, Beijing 100049, China} \maketitle

\begin{abstract}
We speculate that the mass spectrum of three neutrinos might have
a normal hierarchy as that of three charged leptons or that of
three up-type (or down-type) quarks. In this spirit, we propose a
novel parametrization of the $3\times 3$ lepton flavor mixing
matrix. Its mixing angles $\theta^{}_l$ and $\theta^{}_\nu$ can be
related to the mass ratios $m^{}_e/m^{}_\mu$ and $m^{}_1/m^{}_2$
in a specific texture of lepton mass matrices with vanishing (1,1)
elements: $\tan\theta^{}_l = \sqrt{m^{}_e/m^{}_\mu}~$ and
$\tan\theta^{}_\nu = \sqrt{m^{}_1/m^{}_2}~$. The latter relation,
together with solar and atmospheric neutrino oscillation data,
predicts 0.0030 eV $\lesssim m^{}_1 \lesssim$ 0.0073 eV, 0.009 eV
$\lesssim m^{}_2 \lesssim $ 0.012 eV and 0.042 eV $\lesssim m^{}_3
\lesssim$ 0.058 eV. The smallest neutrino mixing angle is found to
be $\theta^{}_{13} \approx \theta^{}_l/\sqrt{2} \approx 3^\circ$,
which is experimentally accessible in the near future.
\end{abstract}

\pacs{PACS number(s): 12.15.Ff, 12.10.Kt}

\newpage

\framebox{\Large\bf 1} ~ The observed pattern of quark flavor
mixing \cite{PDG} remains one of the major puzzles in particle
physics. On account of the hierarchical structures of up- and
down-quark Yukawa couplings in the standard model, a natural and
useful representation of the Cabibbo-Kobayashi-Maskawa (CKM) quark
mixing matrix reads as follows \cite{FX97}:
\begin{eqnarray}
V & = & \left ( \matrix{ c^{}_{\rm u}       & s^{}_{\rm u}     & 0
\cr -s^{}_{\rm u}      & c^{}_{\rm u}     & 0 \cr 0       & 0 & 1
\cr } \right )  \left ( \matrix{ e^{-{\rm i}\varphi}     & 0 & 0
\cr 0       & c     & s \cr 0       & -s    & c \cr } \right )
\left ( \matrix{ c^{}_{\rm d}       & -s^{}_{\rm d}    & 0 \cr
s^{}_{\rm d} & c^{}_{\rm d}     & 0 \cr
0       & 0     & 1 \cr } \right )  \nonumber \\ \nonumber \\
& = & \left ( \matrix{ s^{}_{\rm u} s^{}_{\rm d} c + c^{}_{\rm u}
c^{}_{\rm d} e^{-{\rm i}\varphi} & s^{}_{\rm u} c^{}_{\rm d} c -
c^{}_{\rm u} s^{}_{\rm d} e^{-{\rm i}\varphi} & s^{}_{\rm u} s \cr
c^{}_{\rm u} s^{}_{\rm d} c - s^{}_{\rm u} c^{}_{\rm d} e^{-{\rm
i}\varphi} & c^{}_{\rm u} c^{}_{\rm d} c + s^{}_{\rm u} s^{}_{\rm
d} e^{-{\rm i}\varphi} & c^{}_{\rm u} s \cr - s^{}_{\rm d} s   & -
c^{}_{\rm d} s & c \cr } \right ) \; ,
\end{eqnarray}
where $s^{}_{\rm u} \equiv \sin\vartheta^{}_{\rm u}$, $s^{}_{\rm
d} \equiv \sin\vartheta^{}_{\rm d}$, $c\equiv \cos\vartheta$, {\it
etc}. This parametrization arises automatically from the quark
mass matrices, if the hierarchy of quark masses and their chiral
evolution are taken into consideration. It is not only convenient
for the description of heavy flavor physics (e.g.,
$\tan\vartheta^{}_{\rm u} = | V^{}_{ub} /V^{}_{cb} |$,
$\tan\vartheta^{}_{\rm d} = | V^{}_{td} /V^{}_{ts} |$ and $\sin
\vartheta = \sqrt{|V^{}_{ub}|^2 + |V^{}_{cb}|^2} ~$ hold exactly),
but also suggestive for the understanding of flavor dynamics. For
instance, a variety of models of quark mass matrices predict
\cite{FX99}
\begin{eqnarray}
\tan \vartheta^{}_{\rm u} & = & \sqrt{\frac{m^{}_u}{m^{}_c}} \;\;
,
\nonumber \\
\tan \vartheta^{}_{\rm d} & = & \sqrt{\frac{m^{}_d}{m^{}_s}} \;\;
,
\end{eqnarray}
which are very close to reality and might be exact. The merit of
Eq. (2) is that it provides a phenomenological bridge between the
smallness of quark mixing angles and the hierarchy of quark
masses.

A question is whether the phenomenon of lepton flavor
mixing, which has recently been verified by the solar
\cite{SNO}, atmospheric \cite{SK}, reactor \cite{KM} and
accelerator \cite{K2K} neutrino oscillation experiments, can be
described in a way analogous to Eq. (1). The answer is certainly
affirmative. If neutrinos are Majorana particles, the $3\times 3$
lepton flavor mixing matrix can be expressed as ${\bf V} = UP$,
where
\begin{eqnarray}
U & = & \left ( \matrix{ c^{}_l & s^{}_l   & 0 \cr -s^{}_l    &
c^{}_l   & 0 \cr 0   & 0 & 1 \cr } \right )  \left ( \matrix{
e^{-i\phi}  & 0 & 0 \cr 0   & c & s \cr 0   & -s    & c \cr }
\right )  \left ( \matrix{ c^{}_{\nu} & -s^{}_{\nu}  & 0 \cr
s^{}_{\nu} & c^{}_{\nu}   & 0 \cr
0   & 0 & 1 \cr } \right )  \nonumber \\ \nonumber \\
& = & \left ( \matrix{ s^{}_l s^{}_{\nu} c + c^{}_l c^{}_{\nu}
e^{-i\phi} & s^{}_l c^{}_{\nu} c - c^{}_l s^{}_{\nu} e^{-i\phi} &
s^{}_l s \cr c^{}_l s^{}_{\nu} c - s^{}_l c^{}_{\nu} e^{-i\phi} &
c^{}_l c^{}_{\nu} c + s^{}_l s^{}_{\nu} e^{-i\phi} & c^{}_l s \cr
- s^{}_{\nu} s   & - c^{}_{\nu} s   & c \cr } \right ) \;
\end{eqnarray}
with $c^{}_l \equiv \cos\theta^{}_l$, $c^{}_\nu \equiv
\cos\theta^{}_\nu$, $s \equiv \sin\theta$, and so on; and $P =
{\rm Diag} \left \{e^{i\rho}, e^{i\sigma}, 1 \right \}$ denotes
the Majorana phase matrix. In the approximation that solar and
atmospheric neutrino oscillations are nearly decoupled, the three
mixing angles of $U$ can simply be related to those of solar
\cite{SNO}, atmospheric \cite{SK} and CHOOZ \cite{KM} neutrino
oscillations:
\begin{equation}
\theta^{}_{12} \; \approx \; \theta^{}_\nu \; , ~~~~
\theta^{}_{23} \; \approx \; \theta \; , ~~~~ \theta^{}_{13} \;
\approx \; \theta^{}_l \sin\theta \; .
\end{equation}
This novel parametrization is therefore a convenient option to
describe the present neutrino oscillation data. It is more
convenient than the so-called ``standard" parametrization
\cite{PDG}, when they are applied to deriving the one-loop
renormalization-group equations of relevant neutrino mixing and
CP-violating parameters \cite{Xing05}.

The main purpose of this paper is to explore a possible
relationship between the neutrino mixing pattern and the lepton
mass hierarchy. At first, we speculate that the neutrino mass
spectrum might have a normal hierarchy. Such a hierarchy is
certainly weaker than the mass hierarchy of three charged leptons.
Then we conjecture that the neutrino mixing angles $\theta^{}_l$
and $\theta^{}_\nu$ could be related to the lepton mass ratios
$m^{}_e/m^{}_\mu$ and $m^{}_1/m^{}_2$ in a way similar to Eq. (2).
This hypothesis allows us to determine the magnitudes of three
neutrino masses from current experimental data. It actually
implies an interesting texture of lepton mass matrices with
vanishing (1,1) elements, which can be reconstructed in terms of
the lepton mixing parameters. We find a simple but instructive
possibility to decompose $U$, such that the resultant
charged-lepton and neutrino mass matrices take a parallel form.
The implications of this lepton mass texture will also be
discussed.

\vspace{0.3cm}

\framebox{\Large\bf 2} ~ First of all, let us emphasize that three
mixing angles of $U$ have direct meanings. The angle $\theta$
describes the flavor mixing between the second and third lepton
families. In comparison, $\theta^{}_l$ primarily describes the
$e$-$\mu$ mixing in the charged-lepton sector, and $\theta^{}_\nu$
is essentially relevant to the $\nu^{}_e$-$\nu^{}_\mu$ mixing in
the neutrino sector. In analogy to the quark sector, it makes
sense to conjecture
\begin{eqnarray}
\tan\theta^{}_l & = & \sqrt{\frac{m^{}_e}{m^{}_\mu}} ~ \; ,
\nonumber \\
\tan\theta^{}_\nu & = & \sqrt{\frac{m^{}_1}{m^{}_2}} ~ \; ,
\end{eqnarray}
at least as a leading-order approximation for the lepton mass
models. One can see that Eq. (5) is just in analogy with the
empirically successful Eq. (2) for quark mixing. The possibility
to obtain Eq. (5) from a specific texture of lepton mass matrices
will be discussed later.

The consequences of Eq. (5) are quite non-trivial. Because of
$m^{}_e/m^{}_\mu \approx 0.00484$ \cite{PDG}, we immediately
obtain $\theta^{}_l \approx 4^\circ$ from Eq. (5). This result
indicates that $\theta^{}_{13} \approx \theta^{}_l/\sqrt{2}
\approx 3^\circ$ for $\theta \approx 45^\circ$, as one can see
from Eq. (4). Such a small $\theta^{}_{13}$ is certainly
consistent with current experimental data, $0^\circ \leq
\theta^{}_{13} \leq 10^\circ$ at the $99\%$ confidence level
\cite{Vissani}. Indeed, $\theta^{}_{13} \sim 3^\circ$ is
equivalent to $\sin^2 2\theta^{}_{13} \sim 0.01$, which is
accessible in a few planned reactor neutrino oscillation
experiments (e.g., the Daya-Bay experiment \cite{Wang}). The very
interesting point is that the smallness of $\theta^{}_l$ (or
$\theta^{}_{13}$) is attributed to the strong mass hierarchy of
$e$ and $\mu$ in our conjecture. It is therefore natural and
suggestive for model building.

On the other hand, the observed value $\theta^{}_\nu \approx
33^\circ$ \cite{Vissani} together with Eq. (5) implies that the
mass hierarchy of $\nu^{}_1$ and $\nu^{}_2$ (the mass eigenstates
of $\nu^{}_e$ and $\nu^{}_\mu$) must be rather weak. In
particular, Eq. (5) is qualitatively consistent with the
experimentally-established facts $m^{}_1 < m^{}_2$ and
$\theta^{}_\nu < 45^\circ$ (or equivalently $\theta^{}_{12} <
45^\circ$). Note that solar and atmospheric neutrino oscillations
are associated with the neutrino mass-squared differences $\Delta
m^2_{21} \equiv m^2_2 - m^2_1$ and $\Delta m^2_{32} \equiv m^2_3 -
m^2_2$, respectively. Note also that $\sin^2 \theta^{}_\nu =
m^{}_1 /( m^{}_1 + m^{}_2)$ and $\cos^2 \theta^{}_\nu = m^{}_2 /(
m^{}_1 + m^{}_2)$ hold as a consequence of Eq. (5). Then we arrive
at
\begin{eqnarray}
m^2_1 & = & \frac{\sin^4\theta^{}_\nu}{\cos 2\theta^{}_\nu} \Delta
m^2_{21} \; ,
\nonumber \\
m^2_2 & = & \frac{\cos^4\theta^{}_\nu}{\cos 2\theta^{}_\nu} \Delta
m^2_{21} \; ,
\nonumber \\
m^2_3 & = & \frac{\cos^4\theta^{}_\nu}{\cos 2\theta^{}_\nu} \Delta
m^2_{21} + \Delta m^2_{32} \; .
\end{eqnarray}
Current experimental data yield $\Delta m^2_{21} \approx 8 \times
10^{-5} ~ {\rm eV}^2$ and $|\Delta m^2_{32}| \approx 2.5 \times
10^{-3} ~ {\rm eV}^2$ \cite{Vissani}. Because of
$\cos^4\theta^{}_\nu/\cos 2\theta^{}_\nu \approx 1.2$ for
$\theta^{}_\nu \approx 33^\circ$, the positiveness of $m^2_3$
forbids the possibility of $\Delta m^2_{32} <0$. Then we obtain
$m^{}_1 \approx 0.0041$ eV, $m^{}_2 \approx 0.0097$ eV and $m^{}_3
\approx 0.051$ eV from Eq. (6). Two independent neutrino mass
ratios are $m^{}_1/m^{}_2 \approx 0.42$ and $m^{}_2/m^{}_3 \approx
0.19$. It becomes clear that three neutrino masses have a normal
but relatively weak hierarchy, namely, $m^{}_1 : m^{}_2 : m^{}_3
\approx 1 : 2.4 : 12.5$ as a typical result.

The numerical dependence of two neutrino masses $m^{}_1$ and
$m^{}_2$ on the mixing angle $\theta^{}_\nu$ is explicitly
illustrated in Fig. 1, where $\Delta m^2_{21} = (7.2 \cdots 8.9)
\times 10^{-5} ~{\rm eV}^2$ and $30^\circ \leq \theta^{}_\nu \leq
38^\circ$ (at the $99\%$ confidence level \cite{Vissani}) have
typically  been input. One can see that $m^{}_1$ may change from
0.0030 eV to 0.0073 eV when $\theta^{}_\nu$ increases from
$30^\circ$ to $38^\circ$. In comparison, the allowed range of
$m^{}_2$ is somehow narrower: 0.009 eV $\lesssim m^{}_2 \lesssim $
0.012 eV. Note that $m^{}_3$ is essentially insensitive to
$\theta^{}_\nu$, because $\Delta m^2_{21} \ll \Delta m^2_{32}$ and
$m^{}_3 \approx \sqrt{\Delta m^2_{32}}~$ is a very good
approximation. Given $\Delta m^2_{32} = (1.7\cdots 3.3) \times
10^{-3} ~{\rm eV}^2$ at the $99\%$ confidence level
\cite{Vissani}, we immediately obtain 0.042 eV $\lesssim m^{}_3
\lesssim$ 0.058 eV from Eq. (6).

\vspace{0.3cm}

\framebox{\Large\bf 3} ~ If neutrinos are Majorana particles,
their (effective) mass matrix $M^{}_\nu$ must be symmetric.
Without loss of generality, one may redefine the right-handed
charged-lepton fields to make the charged-lepton mass matrix
$M^{}_l$ to be symmetric too. In this basis, $M^{}_l$ and
$M^{}_\nu$ can be diagonalized by the following transformations:
\begin{eqnarray}
O^\dagger_l M^{}_l O^*_l & = & \left ( \matrix{ \lambda^{}_e & 0 &
0 \cr 0   & \lambda^{}_\mu & 0 \cr 0   & 0 & \lambda^{}_\tau \cr}
\right ) \; ,
\nonumber \\
O^\dagger_\nu M^{}_\nu O^*_\nu & = & \left ( \matrix{ \lambda^{}_1
& 0 & 0 \cr 0   & \lambda^{}_2 & 0 \cr 0   & 0 & \lambda^{}_3 \cr}
\right ) \; ,
\end{eqnarray}
where $O^{}_l$ and $O^{}_\nu$ are unitary matrices,
$\lambda^{}_\alpha$ (for $\alpha = e,\mu,\tau$) and $\lambda^{}_i$
(for $i=1,2,3$) stand respectively for the mass eigenvalues of
charged leptons and neutrinos. Note that $|\lambda^{}_\alpha| =
m^{}_\alpha$ and $|\lambda^{}_i| = m^{}_i$ hold. Alternatively,
one may absorb the negative signs of $\lambda^{}_\alpha$ or
$\lambda^{}_i$ into $O^{}_l$ or $O^{}_\nu$. Note also that the
signs of $\lambda^{}_\alpha$ has no physical significance, but
those of $\lambda^{}_i$ can affect the effective mass term of the
neutrinoless double-beta decay
\begin{equation}
\langle m\rangle^{}_{ee} \; =\; \left | \sum^3_{i=1} \left (
m^{}_i {\bf V}^2_{ei} \right ) \right | \; .
\end{equation}
Because the neutrino mass spectrum has a normal hierarchy in our
scenario, the magnitude of $\langle m\rangle^{}_{ee}$ is strongly
suppressed. It is in general expected that $\langle
m\rangle^{}_{ee} \lesssim {\cal O}(10^{-3})$ eV holds, as a direct
consequence of $m^{}_3 \sim 0.05$ eV and $\theta^{}_{13} \leq
10^\circ$ (or $|{\bf V}^{}_{e3}| \leq 0.17$). This expectation is
actually independent of the details of a specific neutrino mass
model, only if it predicts a normal neutrino mass hierarchy.

It has been mentioned above that the lepton flavor mixing matrix
$\bf V$ is in general written as ${\bf V} = UP$. For simplicity,
we take $U = O^\dagger_l O^{}_\nu$ and attribute the probably
negative signs of $\lambda^{}_i$ into $P$. To reproduce the
parametrization of $U$ in Eq. (3), we may simply express $O^{}_l$
and $O^{}_\nu$ as
\begin{eqnarray}
O^{}_l & = & \left ( \matrix{ e^{i\phi^{}_x} & 0 & 0 \cr 0 &
c^{}_x   & s^{}_x e^{i\varphi^{}_x} \cr 0   & -s^{}_x
e^{-i\varphi^{}_x} & c^{}_x \cr} \right ) \left ( \matrix{ c^{}_l
& -s^{}_l & 0 \cr s^{}_l & c^{}_l & 0 \cr 0 & 0 & 1 \cr} \right )
\; ,
\nonumber \\
O^{}_\nu & = & \left ( \matrix{ e^{i\phi^{}_y} & 0 & 0 \cr 0 &
c^{}_y & s^{}_y e^{i\varphi^{}_y} \cr 0   & -s^{}_y
e^{-i\varphi^{}_y} & c^{}_y \cr} \right ) \left ( \matrix{
c^{}_\nu & -s^{}_\nu & 0 \cr s^{}_\nu & c^{}_\nu & 0 \cr 0   & 0 &
1 \cr} \right ) \; ,
\end{eqnarray}
where $c^{}_x \equiv \cos\theta^{}_x$, $s^{}_y \equiv
\sin\theta^{}_y$, and so on. We find that Eq. (3) can be
reproduced from $U = O^\dagger_l O^{}_\nu$ in four special cases:
\begin{itemize}
\item       Case A: $\varphi^{}_x =0$ and $\varphi^{}_y = 0$. We
obtain $\theta = \theta^{}_y - \theta^{}_x$ and $\phi = \phi^{}_x
- \phi^{}_y$.

\item       Case B: $\varphi^{}_x =\pi$ and $\varphi^{}_y = 0$. We
obtain $\theta = \theta^{}_x + \theta^{}_y$ and $\phi = \phi^{}_x
- \phi^{}_y$.

\item       Case C: $\varphi^{}_x =0$ and $\varphi^{}_y = \pi$. We
obtain $\theta = - \left (\theta^{}_x + \theta^{}_y \right )$ and
$\phi = \phi^{}_x - \phi^{}_y$.

\item       Case D: $\varphi^{}_x =\pi$ and $\varphi^{}_y = \pi$.
We obtain $\theta = \theta^{}_x - \theta^{}_y$ and $\phi =
\phi^{}_x - \phi^{}_y$.
\end{itemize}
If $\theta^{}_x$ and $\theta^{}_y$ are related to the ratios of
charged-lepton and neutrino masses, then it is likely to determine
the mixing angle $\theta$.

With the help of Eqs. (7) and (9), we obtain the charged-lepton
and neutrino mass matrices as follows:
\begin{eqnarray}
M^{}_l & = & \left ( \matrix{ \left ( \lambda^{}_e c^2_l +
\lambda_\mu s^2_l \right ) e^{2i\phi^{}_x} & \left ( \lambda^{}_e
- \lambda^{}_\mu \right ) c^{}_x c^{}_l s^{}_l e^{i\phi^{}_x} &
\left ( \lambda^{}_\mu - \lambda^{}_e \right ) s^{}_x c^{}_l
s^{}_l e^{i\phi^{}_x} \cr \left ( \lambda^{}_e - \lambda^{}_\mu
\right ) c^{}_x c^{}_l s^{}_l e^{i\phi^{}_x} & \left (
\lambda^{}_e s^2_l + \lambda^{}_\mu c^2_l \right ) c^2_x + s^2_x &
\left ( \lambda^{}_\tau - \lambda^{}_e s^2_l - \lambda^{}_\mu
c^2_l \right ) c^{}_x s^{}_x \cr \left ( \lambda^{}_\mu -
\lambda^{}_e \right ) s^{}_x c^{}_l s^{}_l e^{i\phi^{}_x} & \left
( \lambda^{}_\tau - \lambda^{}_e s^2_l - \lambda^{}_\mu c^2_l
\right ) c^{}_x s^{}_x & \left ( \lambda^{}_e s^2_l +
\lambda^{}_\mu c^2_l \right ) s^2_x + c^2_x \cr} \right ) \; ,
\nonumber \\
M^{}_\nu & = & \left ( \matrix{ \left ( \lambda^{}_1 c^2_\nu +
\lambda_2 s^2_\nu \right ) e^{2i\phi^{}_y} & \left ( \lambda^{}_1
- \lambda^{}_2 \right ) c^{}_y c^{}_\nu s^{}_\nu e^{i\phi^{}_y} &
\left ( \lambda^{}_2 - \lambda^{}_1 \right ) s^{}_y c^{}_\nu
s^{}_\nu e^{i\phi^{}_y} \cr \left ( \lambda^{}_1 - \lambda^{}_2
\right ) c^{}_y c^{}_\nu s^{}_\nu e^{i\phi^{}_y} & \left (
\lambda^{}_1 s^2_\nu + \lambda^{}_2 c^2_\nu \right ) c^2_y + s^2_y
& \left ( \lambda^{}_3 - \lambda^{}_1 s^2_\nu - \lambda^{}_2
c^2_\nu \right ) c^{}_y s^{}_y \cr \left ( \lambda^{}_2 -
\lambda^{}_1 \right ) s^{}_y c^{}_\nu s^{}_\nu e^{i\phi^{}_y} & \;
\left ( \lambda^{}_3 - \lambda^{}_1 s^2_\nu - \lambda^{}_2 c^2_\nu
\right ) c^{}_y s^{}_y \; & \left ( \lambda^{}_1 s^2_\nu +
\lambda^{}_2 c^2_\nu \right ) s^2_y + c^2_y \cr} \right ) \; .
\end{eqnarray}
Setting $\left (M^{}_l \right )_{11} = \left (M^{}_\nu \right
)_{11} = 0$ and taking the opposite signs for $\lambda^{}_e$ (or
$\lambda^{}_1$) and $\lambda^{}_\mu$ (or $\lambda^{}_2$), we
immediately arrive at
\begin{eqnarray}
\tan\theta^{}_l & = & \sqrt{- \frac{\lambda^{}_e}{\lambda^{}_\mu}}
\; =\; \sqrt{\frac{m^{}_e}{m^{}_\mu}} \;\; ,
\nonumber \\
\tan\theta^{}_\nu & = & \sqrt{- \frac{\lambda^{}_1}{\lambda^{}_2}}
\; =\; \sqrt{\frac{m^{}_1}{m^{}_2}} \;\; .
\end{eqnarray}
We see that the conjecture made in Eq. (5) can be regarded as a
consequence of the condition $\left (M^{}_l \right )_{11} = \left
(M^{}_\nu \right )_{11} = 0$ which is imposed on the textures of
$M^{}_l$ and $M^{}_\nu$ in Eq. (11). The parallel form and texture
zeros of $M^{}_l$ and $M^{}_\nu$ imply that they might originate
from the same flavor dynamics in an underlying theory.

\vspace{0.3cm}

\framebox{\Large\bf 4} ~
What about the unknown angles
$\theta^{}_x$ and $\theta^{}_y$? Without loss of generality, we
require both of them to lie in the first quadrant. One might
conjecture that $\theta^{}_x$ and $\theta^{}_y$ have the following
relations with the mass ratios $m^{}_\mu/m^{}_\tau$ and
$m^{}_2/m^{}_3$:
\begin{eqnarray}
\sin\theta^{}_x & = & \sqrt{\frac{m^{}_\mu}{m^{}_\tau}} \; ,
\nonumber \\
\sin\theta^{}_y & = & \sqrt{\frac{m^{}_2}{m^{}_3}} \; .
\end{eqnarray}
Then we arrive at $\theta^{}_x \approx 14.1^\circ$ and
$\theta^{}_y \approx 26.1^\circ$ by inputting the experimental
value $m^{}_\mu/m^{}_\tau \approx 0.0594$ \cite{PDG} and the
afore-obtained result $m^{}_2/m^{}_3 \approx 0.19$. It turns out
that only case B mentioned above (i.e., $\varphi^{}_x =\pi$ and
$\varphi^{}_y = 0$) is favored to achieve a large and positive
value for the atmospheric neutrino mixing angle $\theta$: $\theta
= \theta^{}_x + \theta^{}_y \approx 40.2^\circ$. This result,
equivalent to $\sin^2 2\theta^{}_{23} \approx 0.97$, is certainly
consistent with current experimental data. A much larger value of
$\theta$ can be achieved, if we take a larger value for the mass
ratio $m^{}_2/m^{}_3$ in its allowed range.

In Fig. 2, we illustrate the numerical region of $\theta =
\theta^{}_x + \theta^{}_y$ by taking account of Eqs. (6) and (12)
together with current experimental data on $\Delta m^2_{21}$,
$\Delta m^2_{32}$ and $\theta^{}_\nu$ \cite{Vissani}. One can see
that the lower and upper bounds of $\theta$ are about $37.5^\circ$
and $46^\circ$, respectively. In particular, the maximal
atmospheric neutrino mixing (i.e., $\theta \approx 45^\circ$) can
be achieved when $\theta^{}_\nu > 36^\circ$ is input. The simple
correlation between $\theta^{}_\nu$ and $\theta$ shown in Fig. 2
provides a straightforward way to experimentally test the validity
of this phenomenological scenario in the near future.

It seems more difficult to fix the unknown phase parameters
$\phi^{}_x$ and $\phi^{}_y$ in a plausible way, although their
difference $\phi = \phi^{}_x - \phi^{}_y$ is a physical parameter
for CP violation. If CP were a good symmetry in the lepton sector
(i.e., $\phi =0$ held and there were no non-trivial Majorana
phases either), one could simply set $\phi^{}_x = \phi^{}_y =0$.
But it is more likely that there exists large leptonic CP
violation, which might even lead to some observable effects in the
long-baseline neutrino oscillations \cite{X04}. In a few
phenomenological models or Ans$\rm\ddot{a}$tze of lepton mass
matrices (see, e.g., Refs. \cite{FX96} and \cite{Koide}), $\phi
\approx 90^\circ$ or the so-called ``maximal CP violation" has
been discussed. Such discussions might make sense, as current
experimental data on quark flavor mixing and CP violation {\it do}
favor $\varphi \approx 90^\circ$ \cite{Gerard}. It is therefore
interesting to speculate whether $\phi = \varphi = 90^\circ$ can
be realized in a specific model with lepton-quark symmetry.

Given Eq. (12), the textures of $M^{}_l$ and $M^{}_\nu$ in Eq.
(10) can then be illustrated as
\begin{eqnarray}
M^{}_l & \approx & 0.94 m^{}_\tau \left ( \matrix{{\bf 0} & 0.0043
e^{i\phi^{}_x} & 0.0011 e^{i\phi^{}_x} \cr 0.0043 e^{i\phi^{}_x} &
0.0041 & -0.27 \cr 0.0011 e^{i\phi^{}_x} & -0.27
& {\bf 1} \cr} \right ) \; , \nonumber \\
M^{}_\nu & \approx & 0.78 m^{}_3 \left ( \matrix{{\bf 0} & \; 0.14
e^{i\phi^{}_y} \; & -0.094 e^{i\phi^{}_y} \cr 0.14 e^{i\phi^{}_y}
& 0.13 & 0.56 \cr -0.094 e^{i\phi^{}_y} & 0.56 & {\bf 1} \cr}
\right ) \; ,
\end{eqnarray}
where we have taken $\lambda^{}_e = m^{}_e$ (or $\lambda^{}_1 =
m^{}_1$) and $\lambda^{}_\mu = -m^{}_\mu$ (or $\lambda^{}_2 =
-m^{}_2$) by convention and used $m^{}_1/m^{}_2 \approx 0.42$ and
$m^{}_2/m^{}_3 \approx 0.19$ as typical inputs. A normal hierarchy
{\it does} manifest itself in $M^{}_l$ and $M^{}_\nu$. As
expected, the structural hierarchy of $M^{}_\nu$ is relatively
weak. The parallel form and hierarchical textures of $M^{}_l$ and
$M^{}_\nu$, together with their texture zeros, should be helpful
for further model building.

Finally let us point out that Eqs. (5) and (12) or their analogues
can also be reproduced, at least in the leading-order
approximation, from the Fritzsch-type texture of lepton mass
matrices with \cite{Tanimoto} or without \cite{Xing02} the seesaw
mechanism.

\vspace{0.3cm}

\framebox{\Large\bf 5} ~ Starting from the speculation that the
mass spectrum of three neutrinos might have a normal hierarchy in
analogy with that of three charged leptons or three up-type (or
down-type) quarks, we have explored a possible relationship
between the neutrino mixing pattern and the lepton mass matrices.
A novel parametrization, which has proved to be very useful in the
description of quark flavor mixing, is now used to describe the
phenomenon of lepton flavor mixing. Its mixing angles
$\theta^{}_l$ and $\theta^{}_\nu$ are conjectured to relate to the
mass ratios $m^{}_e/m^{}_\mu$ and $m^{}_1/m^{}_2$ in a
straightforward way: $\tan\theta^{}_l = \sqrt{m^{}_e/m^{}_\mu}~$
and $\tan\theta^{}_\nu = \sqrt{m^{}_1/m^{}_2}~$. We find that this
hypothesis allows us to determine the magnitudes of three neutrino
masses and the smallest (CHOOZ) neutrino mixing angle
$\theta^{}_{13}$ from current experimental data. The typical
numerical results are 0.0030 eV $\lesssim m^{}_1 \lesssim$ 0.0073
eV, 0.009 eV $\lesssim m^{}_2 \lesssim $ 0.012 eV and 0.042 eV
$\lesssim m^{}_3 \lesssim$ 0.058 eV, together with $\theta^{}_{13}
\approx \theta^{}_l/\sqrt{2} \approx 3^\circ$. The latter is
experimentally accessible in the near future. We have also pointed
out a simple but instructive possibility to decompose the lepton
mixing matrix, such that the charged-lepton and neutrino mass
matrices can be reconstructed in a parallel form. It turns out
that the texture zeros of their (1,1) elements just lead to the
afore-conjectured expressions of $\theta^{}_l$ and
$\theta^{}_\nu$.

We remark that the structural parallelism, normal hierarchy and
texture zeros could be the useful starting points of view to build
phenomenologically-favored models, in order to deeply understand
the generation of charged-lepton and neutrino masses and the
pattern of their Yukawa-coupling matrices. A similar emphasis has
actually been made for the quark sector \cite{Mei}. Because the
parametrization in Eq. (1) for quark mixing and that in Eq. (3)
for lepton mixing are very analogous to each other, one may even
speculate whether such a novel representation of flavor mixing can
provide a transparent description of the possible lepton-quark
unification in a more fundamental flavor theory.

\vspace{0.5cm}

{\it Acknowledgement:} The work of Z.Z.X. is supported in part by
the National Natural Science Foundation of China.

\begin{figure}[t]
\vspace{2cm}
\epsfig{file=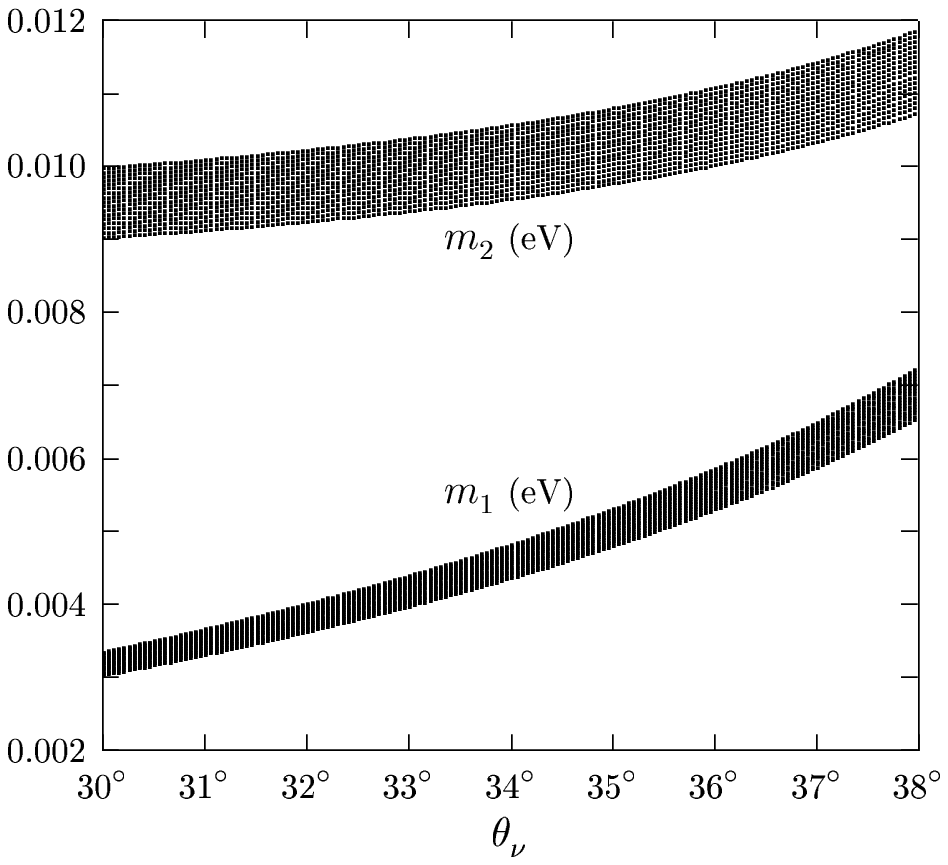, bbllx=1cm, bblly=19cm, bburx=9.8cm, bbury=29cm,%
width=9cm, height=11cm, angle=0, clip=0}
\vspace{1.3cm}
\caption{Changes of two neutrino masses $m^{}_1$ and $m^{}_2$ with
the mixing angle $\theta^{}_\nu$ as predicted by Eq. (6), where
$\Delta m^2_{21} = (7.2\cdots 8.9) \times 10^{-5} ~{\rm eV}^2$ and
$30^\circ \leq \theta^{}_\nu \leq 38^\circ$ [9] have been input.}
\end{figure}

\begin{figure}[t]
\vspace{6cm}
\epsfig{file=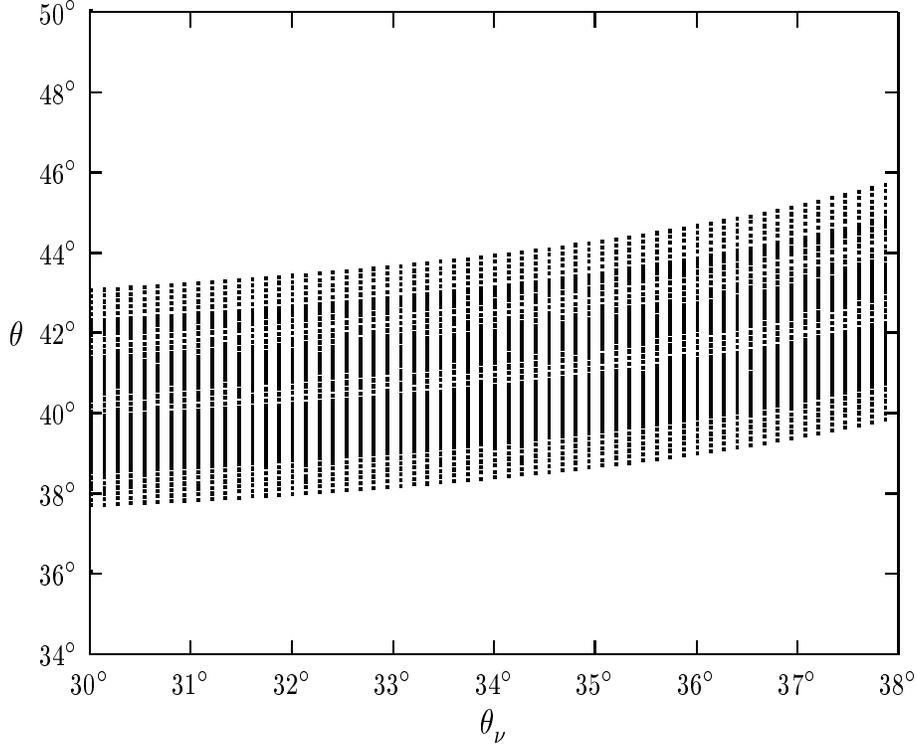, bbllx=1.8cm, bblly=19cm, bburx=11cm, bbury=29cm,%
width=9cm, height=11cm, angle=0, clip=0} \vspace{1.7cm}
\caption{The range of $\theta = \theta^{}_x + \theta^{}_y$
predicted by Eq. (12) and constrained by current data on $\Delta
m^2_{21}$, $\Delta m^2_{32}$ and $\theta^{}_\nu$ in our scenario.
The inputs are $\Delta m^2_{21} = (7.2\cdots 8.9) \times 10^{-5}
~{\rm eV}^2$, $\Delta m^2_{32} = (1.7 \cdots 3.3) \times 10^{-3}
~{\rm eV}^2$ and $30^\circ \leq \theta^{}_\nu \leq 38^\circ$ [9].}
\end{figure}

\end{document}